\documentclass{aa}  

\usepackage{upgreek}
\usepackage{txfonts}
\usepackage{graphicx}
\usepackage{upgreek}
\usepackage{xcolor}
\usepackage{amssymb}

\DeclareMathOperator*{\argmax}{arg\,max}
\DeclareMathOperator*{\dphi}{\frac{\partial}{\partial\phi_o}}
\newcommand{\savg}[1]{\langle#1\rangle}

\newcommand{\edited}[1]{#1}

\begin{document}

        \title{Into nonlinearity and beyond for Zernike-like wavefront sensors}
        
        \author{S.~Y.~Haffert\inst{1}\fnmsep\thanks{NASA Hubble Fellow} }
        
        \institute{University of Arizona, Steward Observatory, Tucson, Arizona, United States\\ \email{shaffert@arizona.edu}}
        \date{Received 18 May 2023; accepted 05 November 2023}
        
        
        \abstract
        {Telescopes like the Extremely Large Telescope (ELT) and the Giant Magellan Telescope (GMT) will be used together with extreme adaptive optics (AO) instruments to directly image Earth-like planets. The AO systems will need to perform at the fundamental limit in order to image Earth twins. A crucial component is the wavefront sensor. Interferometric wavefront sensors, such as the Zernike wavefront sensor (ZWFS), have been shown to perform close to the fundamental sensitivity limit. However, sensitivity comes at the cost of linearity; the ZWFS has strong nonlinear behavior.}
        {The aim of this work is to increase the dynamic range of Zernike-like wavefront sensors by using nonlinear reconstruction algorithms combined with phase sorting interferometry (PSI) and multi-wavelength measurements.}
        {The response of the ZWFS is explored analytically and numerically.}
        {The proposed iterative (non)linear reconstructors reach the machine precision for small aberrations (<0.25 rad rms). Coupling the nonlinear reconstruction algorithm with PSI increases the dynamic range of the ZWFS by a factor of three to about 0.75 rad rms. Adding multiple wavebands doubles the dynamic range again, to 1.4 radians rms.}    
        {The ZWFS is one of the most sensitive wavefront sensors, but has a limited dynamic range. The ZWFS will be an ideal second-stage wavefront sensor if it is combined with the proposed nonlinear reconstruction algorithm.}
        
        
        \maketitle
        
        \section{Introduction}
        Adaptive optics (AO) systems, especially the extreme adaptive optics (ExAO) systems that form part of high-contrast imaging systems, require ever more sensitive and accurate wavefront sensors. The first generation of AO systems used Shack--Hartmann wavefront sensors (SHWFS). The systems with SHWFSs have been very successful; they were the very first to image substellar companions \citep{chauvin2004giant}. However, it was also shown that the SHWFS has relatively poor sensitivity to low-order modes \citep{guyon2005limits}. This is intrinsic to the SHWFS. This sensor uses a micro-lens array (MLA) in the pupil to create an array of spots. A flat wavefront will create a perfect regular pattern. Local slopes in the wavefront map will displace the spots and in turn this displacement signal can be used to reconstruct the incoming wavefront. The fact that the SHWFS measures gradients means that it is more sensitive to high-order modes, because those modes have much stronger derivatives. This is not ideal for direct imaging instruments, which require the best performance at small angular separations, where low-order modes dominate the performance.
        
        The pyramid wavefront sensor (PWFSs) partially solved this issue \citep{ragazzoni1996pupil}. The PWFS uses a pyramidal prism in the focal plane to create four pupils. The PWFS has a very small dynamic range, which can be increased by applying a modulation pattern to the point spread function (PSF) at the cost of sensitivity \citep{ragazzoni1996pupil,verinaud2004nature}. The PWFS is now in use in many systems \citep{pinna2016soul,lozi2019visible,males2022magao}; it has been \edited{chosen as the wavefront sensor} for all new high-contrast imaging systems \citep{kasper2021pcs, males2022conceptual,haffert2022visible,fitzgerald2022planetary} and will be part of upgrades of current systems \citep{boccaletti2022upgrading,perera2022gpi}. The PWFS is a sensor that mixes wavefront slope and phase measurements \citep{verinaud2004nature}. This means that the PWFS still has lower sensitivity to low-order modes, {when modulated,} than a perfect sensor. Several modifications have been suggested to improve the sensitivity of the PWFS. Either by changing the shape of the pyramid prism itself (3-PWFS, Flattened PWFS) or the modulation pattern \citep{fauvarque2017general, schatz2021three, verinaud2023bi}.
        
       \edited{The Zernike wavefront sensor (ZWFS) is} more sensitive than either the SHWFS or the PWFS \citep{guyon2005limits, chambouleyron2021variation}. {The ZWFS \citep{n2013calibration} is a common path interferometer, which means that there are no differential aberrations between the reference beam and the sampled beam. The ZWFS phase mask shifts the phase of the core of the PSF. This maximizes the throughput and enables us to use all available photons.} The phase-shifted core creates the reference beam for the measurement. It is near optimal in terms of robustness against photon noise and read noise \citep{chambouleyron2021variation,chambouleyron2022optimizing}. The sensitivity and accuracy of the ZWFS are demonstrated by the picometer precision in wavefront errors \citep{steeves2020picometer}. However, its response is very nonlinear and not even symmetric around a flat wavefront \citep{n2013calibration}. A nonlinear reconstruction algorithm is required to account for this. This Letter describes a new iterative nonlinear reconstruction algorithm that extends the working range of the ZWFS.
        
        \section{Reconstruction algorithms}
        The basic measurement of the ZWFS is made by the interference of the incoming wavefront and a reference beam. The incoming wavefront is focused onto a focal plane mask and then collimated again to form a pupil. {The reference beam is created by the electric field that passes through the phase dimple on the focal plane mask.} The mathematical description is
        \begin{equation}
                E_{o} = F\{ F\{E_{i}\}  M \}
        .\end{equation}
        Here the input electric field, $E_i$, is filtered by a focal plane mask $M,$ which creates the output electric field $E_o$. The effect of the focal plane mask can be rewritten in a clearer form:

        \begin{equation}
        E_{o} = F\{ F\{E_{i}\} [(M-1)  + 1]\} = E_i + F\{ F\{E_{i}\}  (M-1)\} = E_i + E_r.
        \end{equation}
        This equation shows that the response of any focal plane mask can actually be described by the addition of the input electric field and a (self-created) reference beam. The exact shape of the reference beam depends on the focal plane mask and on the incoming wavefront. The ZWFS uses a top-hat filter of a certain diameter and depth. The optimal size was first thought to be the 50\% encircled energy (EE) diameter at about 1.06$\lambda/D$, which would result in 50\% of the light being part of the test beam and \edited{the other 50\% of the light being part of the reference beam}. Such a diameter creates a reference beam that is larger than the pupil; this is because of diffraction, which then scatters light outside of the geometric pupil \citep{n2013calibration,chambouleyron2021variation}. \citeauthor{chambouleyron2021variation} found that a dot size of $2.0 \lambda/D$ for a clear aperture is better in terms of sensitivity.
 
        Detectors cannot measure the electric field directly, only the intensity is accessible. The intensity pattern of the ZWFS is the classic two-wave interference pattern,
        \begin{equation}
        I = |E_i|^2 + |E_r|^2 + 2|E_i||E_r|\cos{(\phi_i - \phi_r)}.
 \label{eq:main}
        \end{equation}
        Each electric field is parameterized by its amplitude $|E_j|^2=I_j$ and phase $\phi_j$. The phase difference between the input and reference beam can be found by inverting Eq. \ref{eq:main}: 
 \begin{equation}
        \phi_i = \phi_r + \arccos{\left[ \frac{I - I_i - I_r}{2\sqrt{I_iI_r}} \right]}.
 \label{eq:solmain}
        \end{equation}
    This solution is also the \edited{maximum-likelihood estimator (MLE)} for the incoming phase under Poisson noise (bright sources) and Gaussian noise (read noise dominated); see Appendix \ref{app:mle}. However, this equation requires knowledge of the input intensity and the intensity of the reference beam. The incoming pupil intensity $|E_i|^2$ can be measured by taking out the focal plane mask, or simply by offsetting the PSF by several $\lambda/D$. The electric field of the reference beam has to be modeled. The common assumption in the literature \citep{n2013calibration,doelman2019simultaneous} is that the reference beam is created by an aberration-free wavefront scaled by the expected Strehl ratio:
        \begin{equation}
        E_r = \sqrt{S}F^{-1}\{ F\{P\} (M-1)\}.
        \end{equation}
    The aberration-free response is created by filtering the pupil function $P$. However, errors are still introduced due to the aberration-free assumption. The estimate can be improved by iterating the solution {\citep{doelman2019simultaneous}}. Each iteration creates the currently best estimate of the reference beam by using the reconstructed phase from the previous iteration:
        \begin{equation}
        E_{r, n} = \sqrt{S}F^{-1}\{ F\{P \exp{\left(i \phi_{n-1}\right)}\} (M-1)\}.
        \end{equation}
        The Strehl, $S,$ in this method is a correction factor for high-frequency modes that cannot be measured because of the finite number of pixels. The ZWFS uses light from the Airy core for the reference beam. Light from high-spatial frequencies that are not controlled is scrambled because of the atmosphere. This effectively reduces the amount of power in the Airy core, which requires compensation. Simulations in the following section show that the iterative approach always converges to the correct phase if the aberration is within the dynamic range of the sensor.
        
        \subsection{Linear reconstruction}
        The linear reconstructor for small wavefront aberrations can be made by Taylor expanding Eq. \ref{eq:main} around a zero input phase:
    \begin{equation}
        I = I_i + I_r - 2\sqrt{I_iI_r}\sin{(\phi_r)} \phi_i.
        \end{equation}
 
    From this, the linear reconstructor follows as
        \begin{equation}
        \phi \approx \frac{-1}{\sin{(\phi_r)}} \frac{I - I_i - I_r}{2\sqrt{I_iI_r}}. 
        \end{equation}
        The linear approximation can also be used in an iterative approach. The Taylor expansion is then taken around the new reference phase for each iteration. This is already implicitly taken care of by updating the reference electric field.
        
        \subsection{Multistep nonlinear reconstruction}
        The nonlinear reconstructor can account for many of the effects of the interferometric measurement. There is one big limitation. {A sinusoidal signal will have the same value twice over the $[0, 2\pi]$ range, which reduces the reconstructed phase range from both the $\arccos$ and $\arcsin$ to $[0, \pi]$.} This causes {a phase degeneracy} on an interval that is smaller than the optical phase wrapping interval. {Phase degeneracy} can be circumvented using multiple measurements with different reference phases. This approach is called phase-shifting interferometry (PSI) and is the standard approach for metrology systems based on interferometers \citep{groot2011phase}.
 
    There are many different ways to create a different phase response. The phase shift can be applied  either by changing the depth of the ZWFS dot \citep{wallace2011phase,doelman2019simultaneous} or by adding a phase diversity in the incoming wavefront. Modification of the incoming wavefront is a very attractive solution; there is no need for additional hardware because an AO system running in closed-loop automatically creates a time series of different wavefronts. The method that is proposed here is valid for both approaches of phase modulation.

    Each measurement in the time series will have its own reference phase. This leads to a system of equations:
    \begin{equation}
    \label{eq:syseqs}
        \begin{bmatrix}
    I_1\\ \vdots\\ I_n\\ \vdots\\ I_N
    \end{bmatrix}
    =
    \begin{bmatrix}
    I_{1,i} + I_{1,r} +2\sqrt{I_{1,i}I_{1,r}}\cos(\phi_{1,i} - \phi_{1,r}) \\ 
    \vdots \\ 
    I_{n,i} + I_{n,r} +2\sqrt{I_{n,i}I_{n,r}}\cos(\phi_{n,i} - \phi_{n,r}) \\ 
    \vdots \\ 
    I_{N,i} + I_{N,r} +2\sqrt{I_{N,i}I_{N,r}}\cos(\phi_{N,i} - \phi_{N,r}) \\
    \end{bmatrix}
    .\end{equation}
    A least-squares solution to this system can be found by expanding each cosine term in the above equation using the trigonometry identity $\cos(x+y)=\cos(x)\cos(y)-\sin(x)\sin(y)$. Each of the cosine terms can be written as $\cos(\phi_{i,o} - \phi_{i,r}) = \cos(\phi_{o} - \delta\phi_i - \phi_{i,r})$ with $\delta\phi_i$ the phase change at step $i$ with respect to the incoming phase. This can be expanded using the trigonometry identity to
    \begin{equation}
    \cos(\phi_o)\cos(\delta\phi_i + \phi_{i,r})-\sin(\phi_o)\sin(\delta\phi_i + \phi_{i,r}).
    \end{equation}
    The last part of the equation is an inner product between the vector $[\cos(\delta\phi_i + \phi_{i,r}), -\sin(\delta\phi_i + \phi_{i,r}]$ and $[\cos(\phi_o), \sin(\phi_o)]$. The systems of equations in Equation \ref{eq:syseqs} can be rewritten in terms of a single matrix vector multiplication:
    \begin{equation}
        \begin{bmatrix}
    Z_1\\
    \vdots\\ 
    Z_n\\
    \vdots\\
    Z_N
    \end{bmatrix}
    =
    \begin{bmatrix}
    \cos(\delta\phi_1 + \phi_{1,r})& & -\sin(\delta\phi_1 + \phi_{1,r})\\ 
    &\vdots& \\ 
    \cos(\delta\phi_n + \phi_{n,r})& & -\sin(\delta\phi_n + \phi_{n,r})\\ 
    &\vdots& \\ 
    \cos(\delta\phi_N + \phi_{N,r})&  & -\sin(\delta\phi_N + \phi_{N,r})\\
    \end{bmatrix}
    \begin{bmatrix}
    \cos{\phi_o}\\
    \sin{\phi_o}
    \end{bmatrix}
    .\end{equation}
    Here, the variable $Z_n$ is the post-processed intensity,
    \begin{equation}
        Z_n = \frac{I_n - I_{n, i} - I_{n, r}}{2\sqrt{I_{n,i}I_{n,r}}}
    .\end{equation}
    The post-processed intensities are all on the left hand side while the right hand side contains all the modified cosines. The system of equations can be solved by inverting the cosine matrix on the right hand side. The result is that the cosine and sine term of the incoming phase can be fitted.

    This set of equations has to be solved for every pixel in the detector. Recent work has already shown that many small sets of equations can be solved at multiple kHz speeds using GPUs \citep{haffert2021data}. It is therefore not unrealistic to solve such systems of equations upon every iteration during closed-loop control. The final step combines the sine and cosine components to get the full phase,
        \begin{equation}
        \phi = \arctan{\left[\sin{\phi} / \cos{\phi} \right]}.
        \end{equation} {An additional computational burden comes from the optical model propagation to determine the reference beam. The propagation through the ZWFS can be significantly sped up by using a semi-analytical propagation method \citep{soummer2007fast}. This method can reduce the ZWFS propagation to two Tall-and-Skinny matrix-vector multiplications, which can be calculated extremely quickly on a GPU.} 

    \subsection{Multiwavelength phase unwrapping}
    Phase unwrapping is required for phase errors that are larger than 1 $\lambda$ peak to valley. The only way to get around this limitation is to use either multiwavelength measurements \citep{cheng1985multiple} or apply some knowledge about the spatial correlations in order to perform spatial phase unwrapping \citep{ghiglia1998two}. There is a downside to spatial phase unwrapping  in that there must be spatial correlations between adjacent pixels, which might not always be the case (e.g., petal and segments modes for the Extremely Large Telescope (ELT) and the Giant
Magellan Telescope (GMT) \citep{haffert2022phasing, bertrou2022confusion}). Multiwavelength measurements can break the phase wrapping degeneracy because a particular optical path difference (OPD) error ($\delta$) will result in a different amount of phase error for each wavelength ($\phi =2\pi / \lambda \times\delta$). The unwrapped phase can be estimated using the measurements at two wavelengths:
    \begin{equation}
        \hat{\delta} = \frac{1}{2\pi/\lambda_1 - 2\pi/\lambda_2} \mathrm{arg}\left( e^{i(\phi_1 - \phi_2)} \right).
    \end{equation}
    More sophisticated phase unwrapping techniques  also exist that are robust against measurement noise \citep{guo2022robust}.

    \begin{figure}
    \includegraphics[width=\columnwidth]{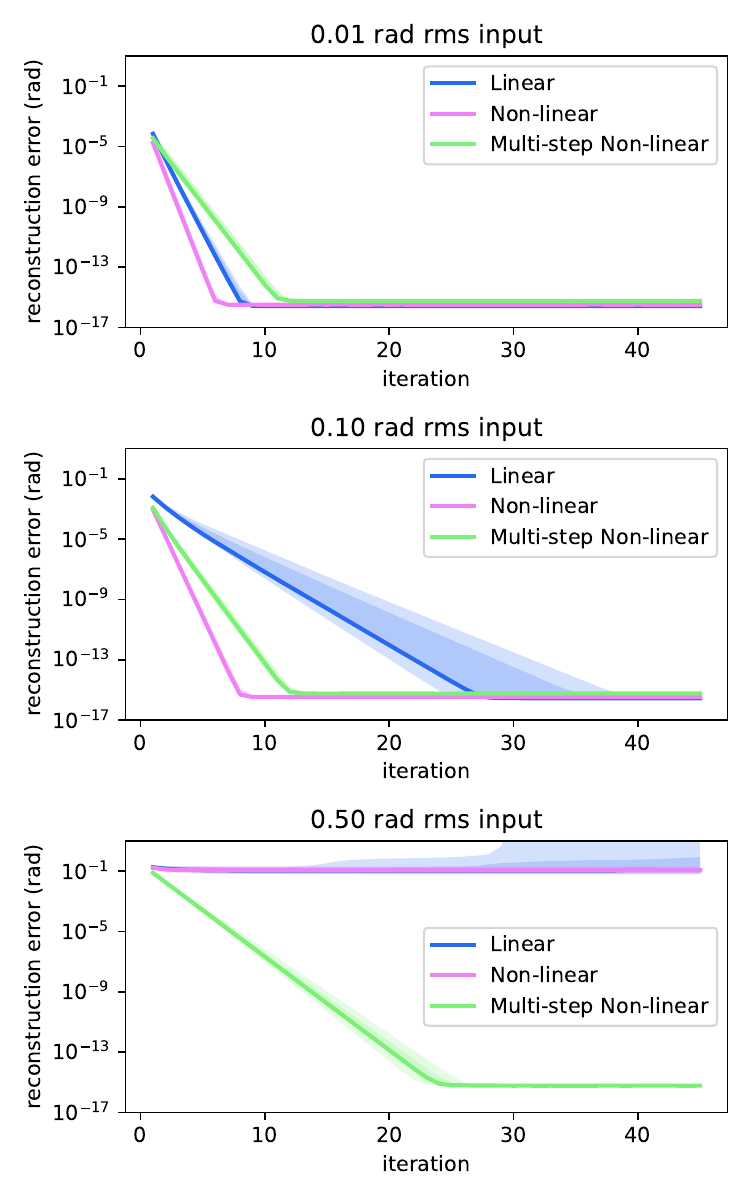}
    \caption{Root mean square of the residuals for different reconstructors as a function of the number of iterations. The different colored lines show the results for the linear (blue), nonlinear (purple), and the multistep (green) reconstructors. The top (0.01 rad rms), middle (0.1 rad rms), and bottom panels (0.5 rad rms) show the result for different aberration strengths. The shaded area represents the 16\% to 84\% percentile, which corresponds to the 1$\sigma$ bounds. The linear and nonlinear reconstructors are able to reach machine precision for small aberrations ($10^{-15}$ relative error for 64bit floats). The nonlinear algorithm converges more quickly than the linear approach. The multistep nonlinear reconstructor converges for all considered aberration strengths. }
    \label{fig:iterations}
    \end{figure}
   
        \section{Simulations}
    The different reconstructors can be compared with one another using numerical simulations. The same setup was used for all simulations presented here, and all of them make use of HCIPy, an open-source object-oriented framework written in Python for performing end-to-end simulations of high-contrast imaging instruments \citep{por2018hcipy}. The standard simulation uses a clear aperture with a telescope diameter of 1m. The exact diameter is not relevant for these sets of simulations because the aberrations are normalized with respect to the diameter of the pupil. The sampling was set to 58 pixels across the pupil. The size of the ZWFS dot was set at 1.5$\lambda/D$ (this size is near optimal sensitivity; see Appendix \ref{app:sensitivity}) and the central wavelength of the simulations is 1 $\mu$m.
    
        \subsection{Required iterations}
    The proposed algorithm uses an iterative approach to update the reference electric field after each estimation of the phase. Upon each iteration, the phase estimate should improve from the updated reference electric field. This behavior is investigated for all three reconstructors (linear, nonlinear, and phase-shifting multistep nonlinear) and for three different aberration strengths: 0.01 rad rms, 0.10 rad rms, and 0.50 rad rms. The aberrations are created from a power-law power spectral density (PSD) with a power-law exponent of -2.5. A total of 31 different trials were performed for each combination of reconstructor and aberration strength to average over statistical effects. The {phase-shifting} multistep approach requires multiple measurements with {a known phase diversity}. {This set of simulations is used to determine the open-loop reconstruction performance. This means that no wavefront control is done. Therefore, a phase diversity needs to be added to the system. The main purpose of the phase diversity is to probe the entire wavefront. The most efficient way to do this is by adding white noise to each actuator, because white noise has a uniform spectral density. Simulations have been performed using other phase probes, such as low-order Zernike modes. Overall the white-noise phase probes showed the best performance, which is why they were used for comparison with the other reconstructors.} Five measurements are taken in sequence with random white noise applied with 0.2 rad rms. The results of this analysis are shown in Figure \ref{fig:iterations}.

    Both the linear and nonlinear algorithm converge to input phase to within machine precision ($10^{-15}$ relative error for 64bit floats) for small aberrations (0.01 and 0.1 rad rms). The nonlinear algorithm does this in fewer iterations than the linear algorithm, even for the smallest aberration (0.01 rad rms). This shows that it is always better to use the nonlinear algorithm than the linear approximation. There is also no benefit to choosing the linear over the nonlinear reconstructor  in terms
of computational complexity. However, both fail when the aberration is stronger, which is shown by the results of the 0.5 rad rms aberrations. The multistep nonlinear algorithm is able to converge for the strongest aberration. The results show that a maximum of 20 steps are required to reach the machine precision limit. This is clearly much more precise than any reconstruction algorithm needs to be, because any amount of measurement noise quickly limits the accuracy.
    
        \subsection{Dynamic range increase}
\begin{figure*}
    \centering
    \includegraphics[width=0.9\textwidth]{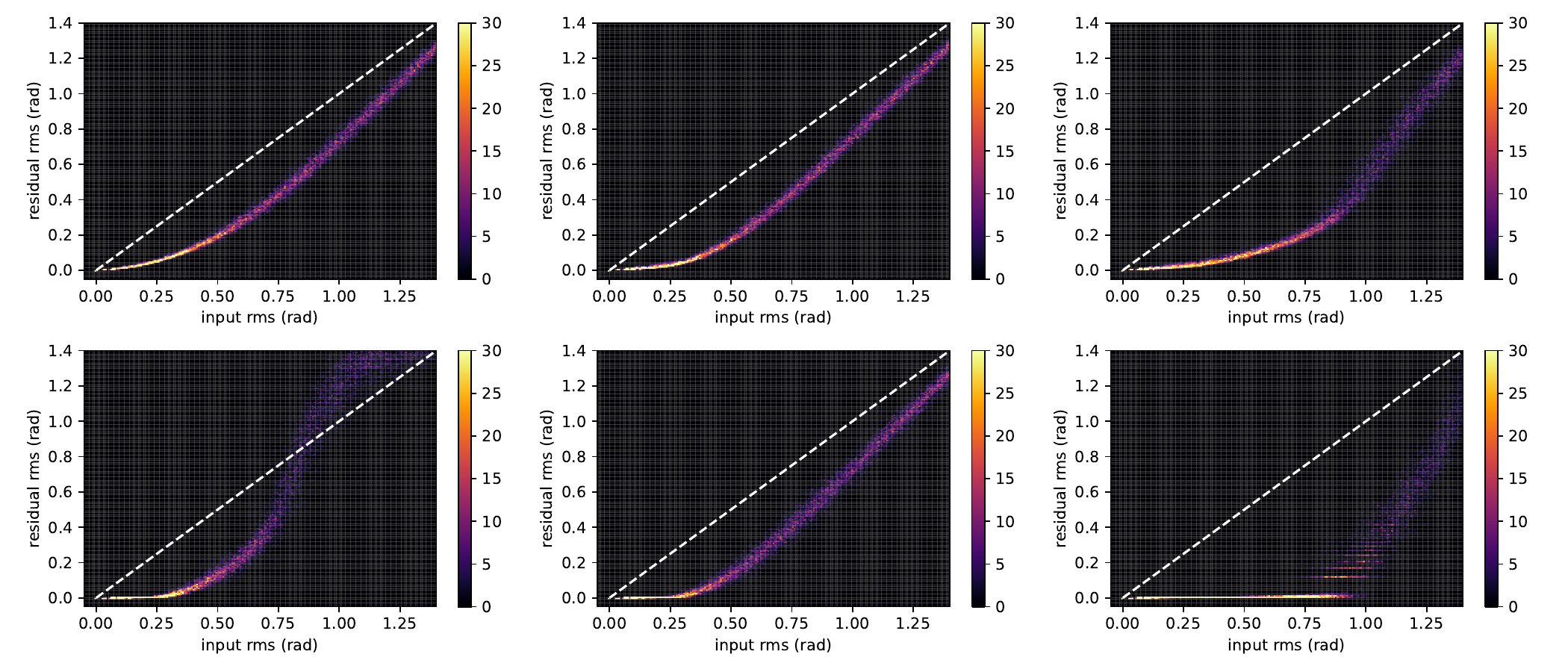}
    \caption{Distribution of the reconstruction errors for different reconstructors (left, linear; middle, nonlinear; right, multistep nonlinear). The top and bottom rows of panels show the error distribution after 1 and after 20 iterations, respectively.}
    \label{fig:reconstruction}
    \end{figure*}

    The iterative algorithm improves the reconstruction quality at all levels of aberration. The increase in performance is explored by stepping through a range of aberrations from 0.001 rad rms to about 1.4 rad rms. Each step has 50 trials, where each trial uses a different realization of the incoming phase. The phase aberration is created using a PSD with a -2.5 exponent in its power law. The incoming aberration is scaled to the required standard deviation after the creation of the phase pattern. The reconstruction algorithms are nonlinear. This means that the actual reconstruction quality not only depends on the total rms but also on how this \edited{power} is distributed across different modes. Nonlinear coupling might lead to situations where certain combination of modes are more difficult to estimate. Therefore, the residual error at each input aberration is a distribution and not a single number. This distribution is shown in Figure \ref{fig:reconstruction}.
    
    The linear and nonlinear reconstructors show similar behavior. After one iteration, the performance of the nonlinear reconstructor is slightly better. However, after 20 iterations, both reach a similar performance. The distributions after 20 iterations are flatter for all reconstructors, indicating a higher reconstruction quality. The linear and nonlinear reconstructors work perfectly until about 0.25 rad rms. After this point, both have decreased performance that roughly scales linearly with the incoming rms. The iterated linear algorithm shows the poorest scaling; its \edited{reconstruction error} increases exponentially at around 0.75 rad rms. The nonlinear algorithm \edited{behaves better and remains stable across all input rms.}. There does not seem to be any difference in its scaling behavior for large input aberrations. The multistep nonlinear reconstructor shows excellent performance at both 1 and 20 iterations; it is superior to the other two reconstructors  in terms of performance. Adding iterations to the multistep algorithm improves its performance even more. The reconstructed error remains flat from around 0 until 0.75 rad rms, beyond which it starts to decrease in performance. The error distribution also starts to become discrete at that point. A closer look at those cases shows that the error is caused by phase wrapping effects. Therefore, there is no possible way to distinguish between those cases and this is therefore the limit of the monochromatic ZWFS.

\begin{figure}
    \includegraphics[width=\columnwidth]{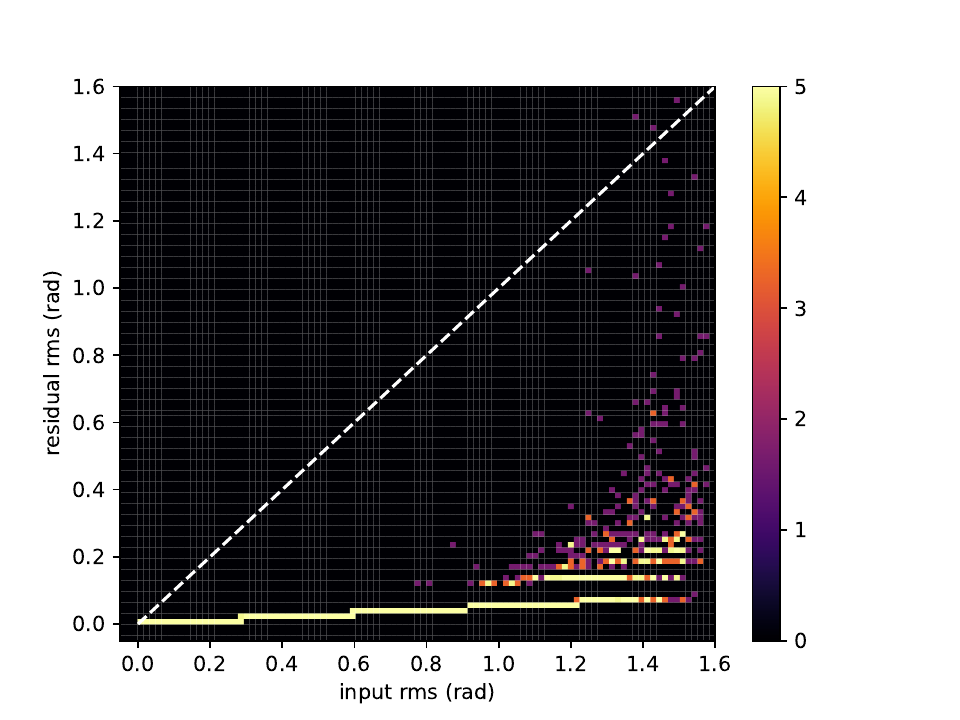}
    \caption{Distribution of the reconstruction errors for the multi-wavelength measurements. The distribution shows a linear trend with increasing input rms, which can be attributed to the limited number of monochromatic iterations. The reconstructor starts to break down around an rms of 1.4 rad, which is 0.23 waves. At this point the phase is starting to wrap multiples times.}
    \label{fig:multi_wave}
    \end{figure}

    Adding multiple wavelengths to the measurements helps to break the degeneracy. With two wavelengths, it is possible to almost double the dynamic range to an rms of 1.4 rad  (Figure \ref{fig:multi_wave}). At that point, the performance of the reconstructor starts to break down because of two issues: The first is that the phase starts to wrap multiple times. This can be solved by using more sophisticated spectral phase unwrappers. The second issue is that the reconstructed monochromatic phase is severely underestimated. This might be an effect of the iterative nature of the reconstructor; it starts with an assumption of a diffraction-limited beam, which creates a reference electric field with a high amplitude. The amplitude becomes much lower if the aberrations are stronger because most of the light will leak around the phase dot. This can only be solved by using a different reconstruction approach, such as machine learning \citep{allan2020deep}.

\begin{figure}
    \includegraphics[width=\columnwidth]{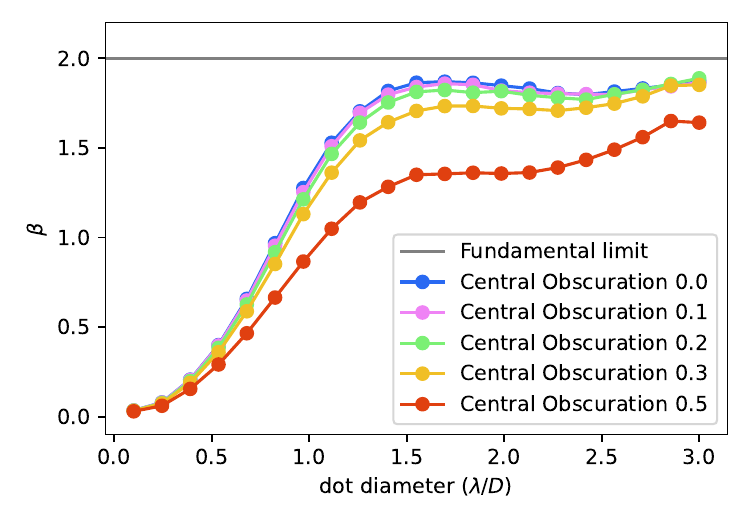}
    \caption{Photon noise sensitivity of the ZWFS for different mask diameters and central obscurations (different colored lines).}
    \label{fig:beta}
    \end{figure}

{
    \subsection{Photon noise sensitivity}
    \label{app:sensitivity}
    When using an iterative algorithm,  we must consider the possible influence of noise and how it propagates through the iterations. The photon noise sensitivity of the nonlinear algorithms is estimated using 
    \begin{equation}
    \label{eq:noise}
    \sigma = \frac{1}{s\sqrt{N}},
    \end{equation}
    where $\sigma$ is the reconstructed phase rms, $s$ is the sensor sensitivity, and $N$ the number of photons that are used for the measurement. In order to average
over statistical noise, 100 different noisy realizations are generated for each number of photons $N$. The phase reconstruction error is then determined as a function of $N$. From this, the sensitivity parameter $s$ can be estimated by fitting Equation \ref{eq:noise} to the data. The results are shown in Figure \ref{fig:beta} for a variety of ZWFS parameters and telescope apertures. The nonlinear iterative algorithm almost reaches the fundamental sensitivity limit for ZWFS with dot diameters of about 1.5 $\lambda/D$. The noise behavior as a function of dot size and central obscuration is very similar to the sensitivity with linear reconstructors \citep{chambouleyron2021variation}. The simulations demonstrate that the iterative nonlinear algorithm does not affect the photon noise sensitivity of the ZWFS. This behavior is also expected because the solution from Equation \ref{eq:main} is the MLE solution. The updates on the reference electric field are minimally impacted by noise because the reference electric field is a low-pass-filtered version of the incoming wavefront \citep{steeves2020picometer}. The low-pass filter reduces the impact of noise.
}
    
        \subsection{Closed-loop simulations}
    {For most AO systems, it is not possible to take multiple measurements with different phase probes.} For AO systems working in closed-loop, another possibility for the PSI \edited{approach} is to make use of the {closed-loop telemetry.} The commands that are sent at every iteration to correct for wavefront aberrations are in fact also phase probes. Therefore, it is possible to \edited{make use of the closed-loop commands for the multistep algorithm}. This makes it a very efficient algorithm, as no iterations are necessary to do the probing. This is similar to the ``Fast and Furious'' focal plane wavefront sensing technique, which uses the control history to solve for degeneracies \citep{keller2012extremely,wilby2018laboratory,bos2020sky}. {The closed-loop control works by keeping track of the last $M$ measurements and control commands that are sent to the DM. The overall phase disturbance is assumed to be static during this period. This is a good assumption for AO systems running at several kHz or for ZWFSs that are used to track quasi-static speckles (e.g., ZELDA on SPHERE). The main challenge is the bookkeeping of the actual phase differences that are applied on the DM. The phase difference is the accumulation of the control signals. If the control signal at timestep $i$ is $v_i$, then the introduced phase difference at timestep $j$ is $\delta \phi_j = v_{j-1} - v_{j-2}$. The time lag of the AO system is assumed to be 1 frame. The full temporal response of the DM has to be taken into account if the dynamics are more complicated than integer delays. The simulations in this work assume a lag of 1 frame.}

\begin{figure}
    \includegraphics[width=\columnwidth]{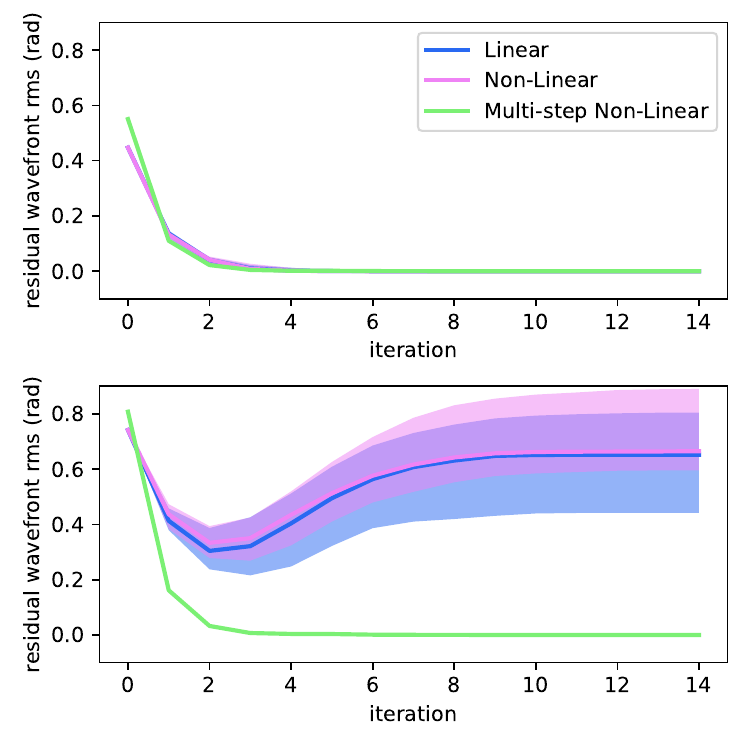}
    \caption{Closed-loop performance as a function of the number of iterations. The solid curve is the median performance across 15 different realizations and the shaded area depicts the 1$\sigma$ boundary. The different colors correspond to the different reconstruction algorithms.}
    \label{fig:closedloop}
    \end{figure}

    Figure \ref{fig:closedloop} shows the wavefront residual rms during closed-loop operations. For small wavefront aberrations, all three algorithms converge to zero wavefront error. However, only the {phase-shifting} multistep algorithm converges for the wavefront aberration starting at 0.8 rad rms. The single-step solutions converge to the wrong phase correction because of the {phase degeneracy} in the reconstruction. The multistep approach is free from this constraint.
 
        \section{Discussion and conclusion}
        This work shows how the phase reconstruction of the Zernike wavefront sensor can be extended using iterative (non)linear reconstructors. The iterative reconstructors increase the accuracy of the ZWFS to the machine precision limit. The ZWFS reconstructor is also extended using PSI techniques. These techniques help the ZWFS to circumvent numerical phase wrapping limits and to increase its dynamic range by a factor of three. The new algorithm, the multistep nonlinear reconstructor, extends the working range of the ZWFS by a factor of four. The main simulations presented here were performed with monochromatic light to demonstrate the iterative and multistep reconstructors. The approach is still applicable for measurements with broadband light, which introduce only a very small error (<0.01 rad rms); see Appendix \ref{app:broadband}. Extending the measurements by using multiple wavebands almost doubles the dynamic range again, allowing the ZWFS to work down to 15\% Strehl. Such measurements are enabled by energy-resolving detectors, such as MKIDS \citep{o2019energy}.

    The ZWFS has increased in popularity over recent years as an ideal wavefront sensor for high-contrast imaging and extreme AO systems. Until now, most instruments only used it as a final stage sensor to sense noncommon path aberrations, because of its small dynamic range \citep{n2016calibration, vigan2019calibration, van2022sky, hours2022harmoni}. However, it is also possible to use the ZWFS as a second stage wavefront sensor. It will most likely still need to operate at IR wavelengths to have a sufficiently large dynamic range. Such an approach is envisioned for the Giant Magellan Adapative Optics eXtreme (GMagAO-X) instrument \citep{haffert2022visible}. A multistage AO system is  planned for this instrument,  where a first-stage AO system is used to control the 4K low-order deformable mirror. This cleans up the wavefront aberrations to such a degree that the ZWFS can be used for the high-order loop that controls the 21.000 actuator deformable mirror \citep{males2022conceptual, close2022optical}. Lab experiments are underway to verify the proposed algorithms for real-life feasibility.
  
        \begin{acknowledgements}
                Support for this work was provided by NASA through the NASA Hubble Fellowship grant \#HST-HF2-51436.001-A awarded by the Space Telescope Science Institute, which is operated by the Association of Universities for Research in Astronomy, Incorporated, under NASA contract NAS5-26555.
        \end{acknowledgements}

        
        \bibliography{references}   
    \bibliographystyle{aa}   
        
\begin{appendix}
\section{Maximum-likelihood derivation}
\label{app:mle}
  The maximum-likelihood solution starts by considering the likelihood. We are considering two cases for the Zernike wavefront sensor; a Gaussian-noise model and a Poisson-noise model. A major advantage to using the Zernike WFS is that it is a direct wavefront sensor. There is no information required from neighbor pixels to reconstruct the phase of a single pixel. Therefore, we can consider all pixels as independent problems, making it easy to find the MLE solution.  
                
                \subsection{Gaussian-noise maximum-likelihood estimator}
  The likelihood, $\mathcal{L}_i$, of a Gaussian-distributed variable $x_i\sim N(\mu_i, \sigma_i)$ is
        
        \begin{equation}
         \mathcal{L}_i = \frac{1}{\sqrt{2\pi\sigma_i^2}}\exp{\left(-\frac{(x_i - \mu_i)^2}{\sigma_i^2}\right)}.
         \label{eq:likelihood}
        \end{equation}
        
        The data point $x_i$ is the measured intensity value and $\sigma_i$ the noise for that data point. The expectation value follows from Eq. \ref{eq:main} as $\mu_i = I_o + I_r + 2\sqrt{I_o I_r}\cos{(\phi_o - \phi_r)}$. The MLE of the phase is found by maximizing the likelihood with respect to the phase. It is possible to directly optimize the likelihood, but it is easier to optimize the logarithm of the likelihood. Maximizing Eq. \ref{eq:likelihood} is the same as finding the maximum of the logarithm of the equation,
        \begin{equation}
            \hat{\phi} = \argmax_{\phi} \mathcal{L}_i = \argmax_{\phi} \log{\mathcal{L}_i}.
        \end{equation}
        The maximum can be found by equating the derivative of the log-likelihood to zero,
        \begin{equation}
        \dphi\log{\mathcal{L}} = -\frac{1}{2} 2\pi\dphi\sigma_i^2 - \dphi\frac{(x_i - \mu_i)^2}{\sigma_i^2} = 0.
        \end{equation}
        Here, only the second term depends on the incoming phase.
        \begin{equation}
        \dphi\log{\mathcal{L}} =  2\frac{(x_i - \mu_i)}{\sigma_i^2}\dphi \mu_i = 0.
        \end{equation}
        The equation shows we have two possible solutions:
        \begin{equation}
        x_i - \mu_i = 0 \vee \dphi \mu_i = 0.
        \end{equation}
        The first equation is just the direct solution of Eq. \ref{eq:main}. The other solution results in
        \begin{equation}
        \sin{(\phi_o - \phi_r)} = 0.
        \end{equation}
        This solution no longer contains the measurement $x_i$  and does not give us any information. Second, the first solution always has a smaller log-likelihood because $x_i - \mu_i$ is perfectly canceled, which is not always the case for the second equation. 
                
        \subsection{Poisson-noise maximum-likelihood estimator}
        The Poisson-noise MLE can be derived in a straightforward manner. The Poisson distribution is parameterized by the rate parameter $\nu$. The corresponding distribution is
        \begin{equation}
        \mathcal{L}_i = \exp(-\nu)\frac{\nu^{x_i}}{x_i!}.
        \end{equation}
        The log-likelihood is
        \begin{equation}
        \log{\mathcal{L}_i} = -\nu -\log{x_i!} + x_i\log{\nu}.
        \end{equation}
        For the ZWFS, the photon rate is set by Eq. \ref{eq:main}. This means that $\nu = I_o + I_r + 2\sqrt{I_o I_r}\cos{(\phi_o - \phi_r)}$. The derivative of the log-likelihood with respect to the incoming phase, $\phi_o$, is
        \begin{equation}
        \dphi \log{\mathcal{L}_i} = -\dphi \nu  + x_i\dphi \log{\nu},
        \end{equation}
        \begin{equation}
         \dphi \log{\mathcal{L}_i} = 2\sqrt{I_oI_r}\sin{\delta\phi} - x_i \frac{2\sqrt{I_oI_r}\sin{\delta\phi}}{I_o + I_r + 2\sqrt{I_oI_r}\cos{\delta\phi}}.
        \end{equation}
        The MLE solution can be found by equating the derivative to zero and solving for $\phi_o$. After some algebra this becomes
        \begin{equation}
          I_o + I_r + 2\sqrt{I_oI_r}\cos{\delta\phi} = x_i.
        \end{equation}
        The MLE solution is just the direct solution of the Eq. \ref{eq:main}.
        
    The derivations in this appendix show that the solution in Equation \ref{eq:solmain} is the MLE solution at all photon rates.

    \subsection{MLE solution for a linear combination of modes}
    The equivariance principle of MLE solutions can be used to find the MLE solution in mode spaces  other than the direct pixel basis that was used in the derivations in the previous section. This principle states that if $\hat{\theta}$ is an MLE of $\theta$, then $\hat{\tau} = g(\hat{\theta})$ is the MLE solution of $g(\theta)$ for any function $g$ \citep{casella2021statistical}. Using this principle, we can find the MLE solution of $\alpha = P\phi_o$ with $\alpha$ the mode coefficients and $P$ the modal projection matrix. The MLE solution of $\phi_o$ is found with Eq. \ref{eq:solmain}, which means that the MLE solution of the mode coefficients is $\hat{\alpha}=P\hat{\phi_o}$. The best way to estimate the modal coefficients is to start by solving in pixel space and then project that phase onto the mode basis.

\section{Broadband effects}
\label{app:broadband}
    \subsection{Broadband reconstruction errors}
    The proposed reconstructors are all based on the monochromatic interference pattern. Real AO systems always have a broad spectral bandwidth in order to capture as many photons as possible. The response will be an average over all wavelengths. The spectral average operator $\langle\cdot\rangle$ is defined as
    \begin{equation}
        \langle f \rangle = \frac{\int_{\lambda_0 - \Delta \lambda/2}^{\lambda_0 + \Delta \lambda/2} f(\lambda) \mathrm{d}\lambda}{\int_{\lambda_0 - \Delta \lambda/2}^{\lambda_0 + \Delta \lambda/2} \mathrm{d}\lambda}.
        \end{equation}

\begin{figure}
    \includegraphics[width=\columnwidth] {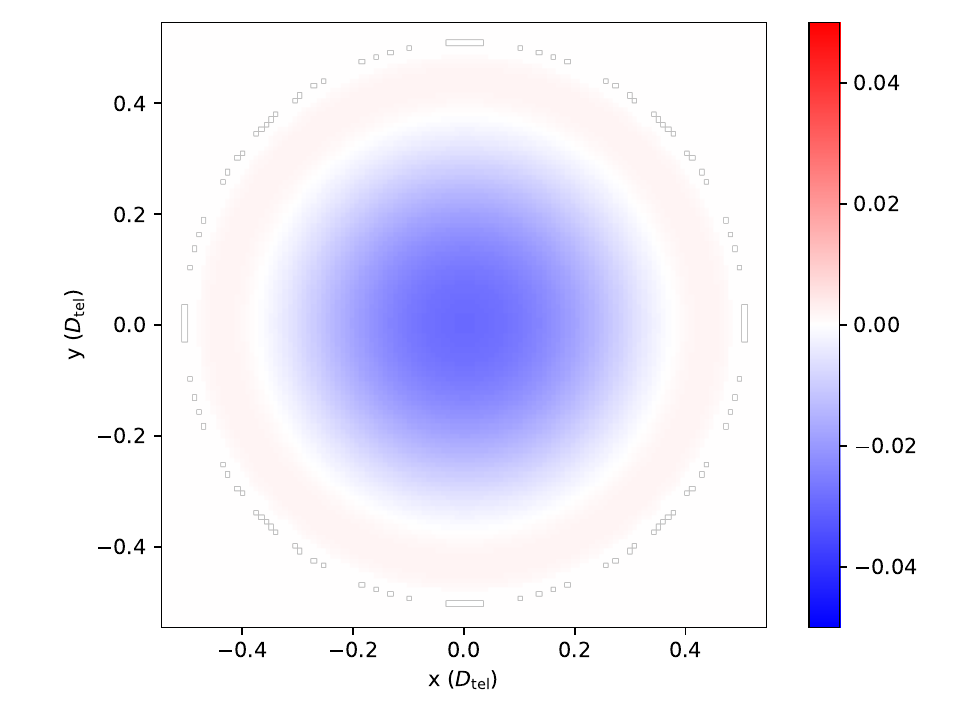}
    \caption{Phase error in radians due to the chromatic change in the reference field.}
    \label{fig:avg_ref}
    \end{figure}

    The wavelength averaged ZWFS response is
    \begin{equation}
        \savg{I} = \savg{I_i} + \savg{I_r} + 2\savg{\sqrt{I_i I_r}\cos{(\phi_i - \phi_r)}}.
    \end{equation}
    The first term is just the incoming pupil, which is nearly achromatic. The only chromatic effects that causes differential amplitude variations in the pupil are due to scintillation. However, such effects are usually very small. The pupil image can be assumed to be achromatic, $\savg{I_i} = I_i$. The second term is the spectral average of the reference electric field. The changes in the reference beam across wavelengths of of low order. The reference field is created by the circular ZWFS mask. This creates an Airy pattern in the pupil. This pupil-plane Airy pattern scales with wavelength due to diffraction, which is the first chromatic effect. The second comes from the incoming beam. The amount of power that passes through the ZWFS dimple also scales with wavelength due to diffraction; lower wavelengths have a smaller PSF. In summary, shorter wavelengths have a more concentrated reference beam with higher power, while longer wavelengths have a more spread out reference beam with less power. These two effects are partially compensated by spectral averaging. Figure \ref{fig:avg_ref} shows the phase error from the change in the reference field averaged over a 50\% bandwidth. The phase error consists of 0.009 rad rms focus error and 0.0036 rad rms of spherical error. This is a very small error and can be safely neglected in the reconstruction.    
    
    The inversion of the measurement equation is complicated by the last term. It is a weighted integral of a cosine, which cannot simply be replaced by a single cosine. The error from the monochromatic assumption can be estimated by making a Taylor expansion of the integral for small bandwidths around the center wavelength $\lambda_0$,

    \begin{equation}
        \int_{\lambda_0 - \Delta \lambda/2}^{\lambda_0 + \Delta \lambda/2}\sqrt{I_i I_r}\cos{\left(\phi_i - \phi_r\right)} + B(\lambda - \lambda_0) + C(\lambda - \lambda_0)^2 \mathrm{d}\lambda.
        \end{equation}
    Here, $B$ and $C$ are the first- and second-order expansion coefficients. The first-order term will cancel if the integral is taken over a symmetric domain around the center wavelength. This will leave us with
    \begin{equation}
        2 \sqrt{I_i I_r} \cos{\left(\phi_i - \phi_r\right)} \Delta \lambda  + \frac{C}{6}\Delta \lambda^3.
        \end{equation}
    The spectral average is normalized by the total bandwidth $\Delta \lambda$. Therefore, the integral term has to be divided by $\Delta \lambda$. Substituting this back into the original broadband equation leaves us with
    \begin{equation}
        \savg{I} = I_i + I_r + 2 \sqrt{I_i I_r} \cos{\left(\phi_i - \phi_r\right)}  + \frac{C}{6}\Delta \lambda^2.
    \end{equation}
    
\begin{figure}
    \includegraphics[width=\columnwidth]{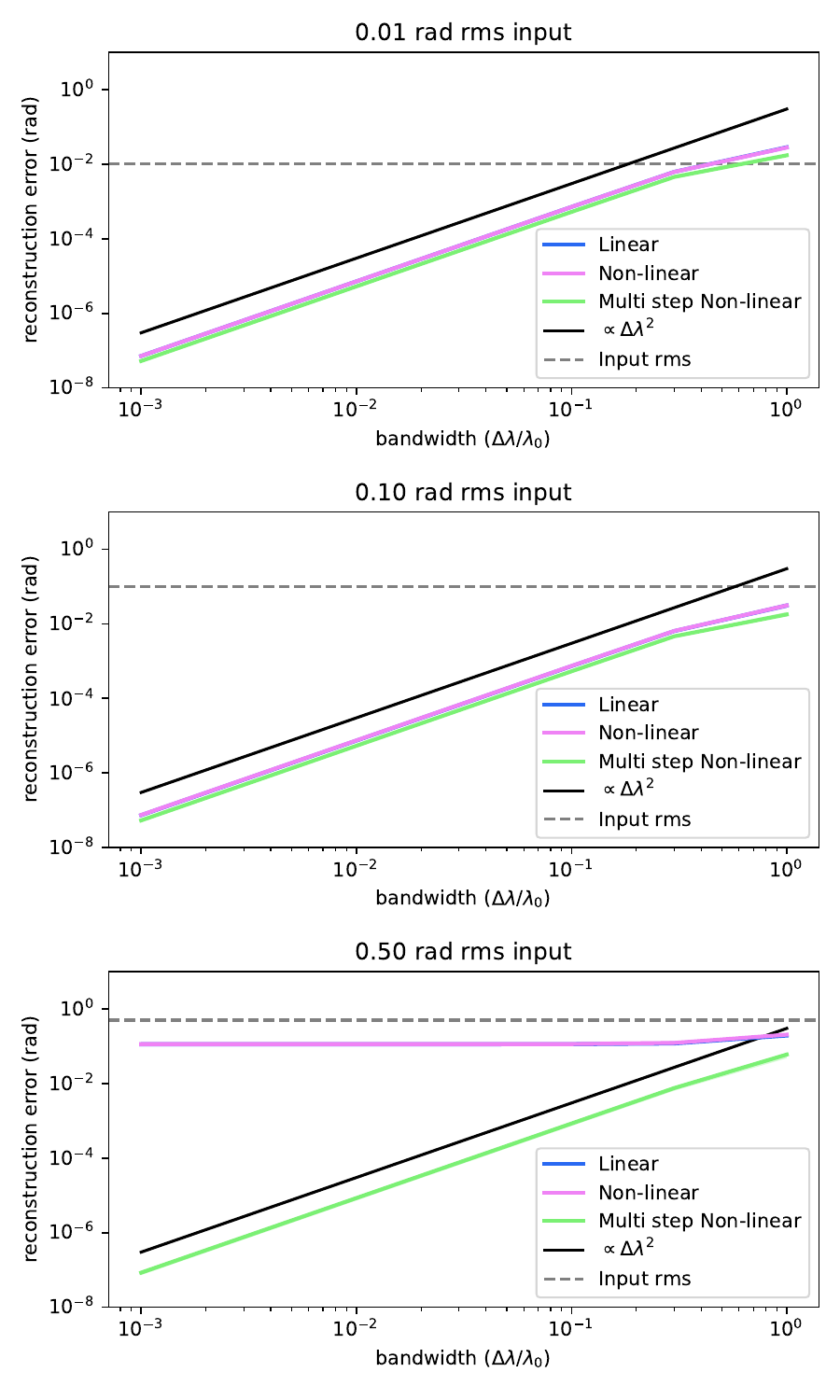}
    \caption{Reconstruction error as a function of bandwidth for the different reconstructors. The top, middle, and bottom panels show the reconstruction error for different input rms values. The top and middle panels show that the reconstruction error grows as the bandwidth squared. The bottom panel shows the results for an rms value that is outside the dynamic range of the single-step algorithms.}
    \label{fig:bandwidth}
    \end{figure}

    This derivation shows that using the monochromatic solution has a second-order error in spectral bandwidth. The second-order behavior is made clear by Figure \ref{fig:bandwidth}, which shows the residual wavefront error rms as a function of spectral bandwidth. The typical bandwidths for wavefront sensors are at the 0.1 to 0.5 level. At such bandwidths, the reconstruction error is roughly 0.01 rad rms. This is substantially lower than any measurement noise. The reconstruction error created by a finite bandwidth will therefore be of no consequence and the monochromatic reconstructor can be used without considering bandwidth effects.
    
    \end{appendix}
    
\end{document}